# A Survey on Parallel Genetic Algorithms for Shop Scheduling Problems

Jia LUO
LAAS-CNRS, Universite de Toulouse, CNRS
Toulouse, France
jluo@laas.fr

Didier EL BAZ
LAAS-CNRS, Universite de Toulouse, CNRS
Toulouse, France
elbaz@laas.fr

*Abstract*—**There have been extensive works dealing with genetic algorithms (GAs) for seeking optimal solutions of shop scheduling problems. Due to the NP hardness, the time cost is always heavy. With the development of high performance computing (HPC) in last decades, the interest has been focused on parallel GAs for shop scheduling problems. In this paper, we present the state of the art with respect to the recent works on solving shop scheduling problems using parallel GAs. It showcases the most representative publications in this field by the categorization of parallel GAs and analyzes their designs based on the frameworks.**

## I. Introduction

The Genetic algorithm (GA) is a stochastic search algorithm based on the principle of natural selection and recombination [1]. It is a kind of evolutionary algorithm and has been successfully applied to solve many optimization problems, e.g., knapsack problems, shop scheduling problems or travelling salesman problems [2][3][4]. Nevertheless, when GAs are applied to more complex and larger problems, the required time to find adequate solutions is increased. Particularly, repeated fitness function evaluation is often the most prohibitive and limiting segment when GAs are hired to find an optimal solution for high-dimensional or multimodal implementations. Meanwhile, GAs also suffer from the problem of a tendency to converge towards local optima rather than the global optimum. Previous works in this area suggest to enlarge population size, increase mutation rate or hire niche penalty in selection to keep the diversity of GAs. However, any of them may raise the complexity of the algorithm and lead to more time consumption.

No doubt, parallel implementation is considered as one of the most promising choices to make GAs work faster. There are different ways of exploiting parallelism in GAs [5]: master-slave models, fine-grained models, island models and hybrid models. The master-slave model is the only one that does not affect the behavior of the algorithm by distributing the evaluation of fitness function to slaves. The fine-grained model works with a large spatially population. The evolution operations are restricted to a small neighborhood with some interactions by overlap structure. The island model divides populations into subpopulations. Subpopulations on the islands are free to converge towards different sub-optima and a migration operator can help mix good features that emerge from the local island. The hybrid model combines any two of the above methods.

The shop scheduling problem is one of the best known combinatorial optimization problems where jobs are assigned to machines at particular times. The use of evolutionary algorithms for shop scheduling problems started around 1980 [6]. There have been a huge number of publications dealing with GAs for shop scheduling problems. Due to the NP hardness, the time cost to obtain an adequate solution by the serial GA is always heavy. With the development of high performance computing (HPC) in last decades, the implementation of parallel GAs to shop scheduling problems has been extensively studied. The purpose of this paper is to give a tutorial survey of recent works on solving scheduling problems in manufacturing systems using parallel GAs.

The rest of this paper is organized as follows. In section 2, a brief introduction about shop scheduling problems and their new integrated factors are given. Section 3 discusses the typical parallel GAs, namely, master-slave models, fine-grained models and islands models, developed for the scheduling problems in manufacturing systems. Section 4 analyzes the frequently used HPC frameworks and their associated parallel GAs design in this domain. Finally, conclusions are stated in section 5.

## II. Shop Scheduling Problems

The shop scheduling problem is a classic optimization problem. One instance of the problem consists of a set of n jobs $J_1, J_2, ..., J_i, ..., J_n$ and a set of o machines $M_1, M_2, ..., M_m, ..., M_o$. Each job $J_i$ comprises a number of g stages $S_1, S_2, ..., S_s, ..., S_g$. The processing time of one step of $J_i$ on a particular machine is denoted as an operation and is abbreviated by (j, s, m). Usually, it is given in advance as $P_{jsm}$ with the release time $R_j$ and the due time $D_j$. Additionally, other required conditions are shown in Table 1.

TABLE I. OTHER REQUIRED CONDITIONS FOR SHOP SCHEDULING PROBLEMS

| NO. | Description |
|---|---|
| 1 | Each operation of a job must be processed by one and only one machine. |
| 2 | Each machine can process no more than one operation at a time. |
| 3 | Each job is available for processing after the release time. |
| 4 | Setup times for job processing and machine assignment times between stages are not taken into consideration. |
| 5 | There is infinite intermediate storage between machines. |

There are three ways to classify the scheduling problem in manufacturing systems by the machine environment, the job characteristics and the optimization criterion [7]. Most of the works concern on the three basic types: a flow-shop, a job-shop and an open-shop. In a flow-shop, each job passes through the machines with the same order whereas a job-shop enables specified jobs have possibly different machine orderings. In an open-shop, there is no particular route imposed on jobs. In addition to theses three types, flexible shops also catch a lot of attentions. It is a combination of a shop scheduling problem and a parallel machine scheduling problem, in which at least one stage consists of several parallel machines [6]. Most of the works considered are the flexible flow shop or the flexible job shop.

When a feasible schedule is given, we can compute for each $J_i$: the completion time $C_j$, the tardiness $T_j = \max\{0, C_j - D_j\}$, and the unit penalty $U_j = 1$ if $C_j > D_j$, otherwise 0. The most common optimality criteria are the minimization of the makespan $C_{max}$, the minimization of the sum of the weighted completion time $\sum w_j C_j$, the minimization of the sum of the weighted tardiness $\sum w_j T_j$, and the minimization of the sum of the weighted unit penalty $\sum w_j U_j$, or any combination among them.

With the development of modern manufacturing, some new factors are integrated into the basic shop scheduling problems, such as energy controlling, dynamic environment and so on. Xu et al. built a discrete-time mixed-integer programming model and a slot-based mixed-integer programming model in [8] to achieve a global optimal solution between the peak power and the traditional production efficiency without any compromise on computing efficiency. Tang et al. [9] adopted a predictive reactive approach based on an improved particle swarm optimization to search for the Pareto optimal solution in dynamic flexible flow shop scheduling problems reducing the energy consumption and the makespan.

Most shop scheduling problems are known as strong NP-hard problems [10]. Many works to solve it by exact methods and meta-heuristic methods have been done. However, this class of problems requires complex and time-consuming solution algorithms. Although the speed of the best supercomputer increases 10 times each 3 or 4 years recently, this increase has only a little influence on the size of solvable problems [11]. Therefore, efforts to coordinate these algorithms with HPC accelerators to solve shop scheduling problems efficiently and effectively are deeply desirable.

III. GENETIC ALGORITHMS WITH SCHEDULING PROBLEMS IN MANUFACTURING SYSTEMS

A. Simple Genetic Algorithms

A simple GA [1] starts with a randomly generated initial population consisting of a set of individuals. An individual is representative by a chromosome. For flow shop problems, a standard chromosome consists of a string of length n, and the i-th gene contains the index of the job at position i [6]. An individual describes a feasible schedule of jobs' sequence on target machines. For job shop problems, there are two ways of chromosome representation: direct way and indirect way. The direct way is similar with the way for flow shop problems: a feasible schedule is directly encoded into the chromosome, whereas the chromosome in the indirect way shows a sequence of dispatching rules for job assignment [12]. As no imposed technological routes of the jobs for open shop problems, both of the encoding approaches for the flow shop and the job shop can be applied in this case. The fitness value of each individual is used to evaluate the current population. It is related to the objective function value of shop scheduling problems at the point represented by a chromosome. Since most common optimality criteria of shop scheduling problems are about minimization. The fitness function FIT(i) of an individual i usually can be transferred as [6]:

$$FIT(i) = max(\bar{F} - F_i(S_i), 0) \qquad (1)$$

where $F_i(S_i)$ denotes the objective function value of a feasible schedule from individual i and $\bar{F}$ states the objective function value of some heuristic solution.

As the value of object function for shop scheduling problems are generally positive, some papers measure the fitness function FIT(i) as:

$$FIT(i) = \frac{1}{F_i(S_i)} \qquad (2)$$

Three GA operations: selection, crossover and mutation, work on these chromosomes to get new search points in a state of space. Usually, individuals are first selected through a fitness-based process. For shop scheduling problems, solutions with larger fitness values are more likely to be selected. Some well-known methods are implemented in this step: the roulette wheel selection, the stochastic universal sampling, the tournament selection and so on [13]. Next, the crossover takes two random individuals kept after selection and exchanges random sub-chromosomes. The classic methods are the n-point crossover and the uniform crossover. Due to particular requirements of different shop scheduling problems, additional steps may be required to repair the illegal offspring caused by the crossover. The mutation then alters some random value within a chromosome. Different from the binary encoding, the mutation for shop scheduling problems works often based on the neighborhoods e.g. shift mutation (insertion neighborhood) or pairwise interchange mutation (swap neighborhood) to respect feasible solutions. Population evaluation is executed after these three steps. Sometimes, an elitist strategy is hired afterwards to keep limited number of individuals with the best fitness values to the next generation. This process repeats until the termination criteria have been satisfied. The full procedure is stated in Table 2.

TABLE II. PSEUDO-CODE OF THE SIMPLE GA

| | |
|---|---|
| 1: | initialize(); |
| 2: | while (termination criteria are not satisfied) do |
| 3: |    Generation++ |
| 4: |    Selection(); |
| 5: |    Crossover(); |
| 6: |    Mutation(); |
| 7: |    FitnessValueEvaluation(); |
| 8: | end while |

## B. Master-Slave Genetic Algorithms

Master-slave GA is known as global parallel GA as well. It keeps a single population as a simple GA that is stored at the master side. In this case, each individual is free to compete and mate with any other. Since the fitness value calculations of individuals are independent without any communication with others, the slaves take care of fitness evaluation in parallel. Data exchange occurs only when sending and receiving tasks between the master and slaves. Obviously, frequent communication overhead offsets some performance gains from slaves' computing. However, as master-slave GA is the easiest parallel model to be implemented and does not assume underlying architecture, it is still very efficient when the evaluation is complex and requires considerable computation. The steps of this parallel model are presented in Table 3.

TABLE III Pseudo-code of the master-slave GA

| | |
|---|---|
| 1: | Initialize(); |
| 2: | while (termination criteria are not satisfied) do |
| 3: | Generation++ |
| 4: | Selection(); |
| 5: | Crossover(); |
| 6: | Mutation(); |
| 7: | Parallel_FitnessValueEvaluation_Individuals(); |
| 9: | end while |

### 1) job shop scheduling problems

AitZai et al. modeled the job shop scheduling problem with blocking using the alternative graph with conjunctive arcs and alternative arcs in [14]. In addition to a parallel branch and bound method, two master-slave GA parallelization methods were also presented. The first one was based on CPU networking with a star network of inter-connected computers. On the opposite, the second one worked on GPU with some memory management respecting to CUDA (Compute Unified Device Architecture) framework, which was a NVIDIA's parallel computing architecture that increased computing performance by harnessing the power of GPU. Numerical tests were performed on a station equipped with CPU×2: Intel Xeon E5620 and GPU: NVIDIA Quadro 2000 01 Go GPU. With a population size 1056 and a limited total execution time 300s, the master-slave GA using GPU could get maximum 15 times more explored solutions than the GA using CPU. Moreover, a related earlier work was introduced by AitZai in [15]. In order to improve the solution of job shop scheduling problems, Somani et al. [16] imposed a topological sorting step to the GA before the fitness value calculation, which was used to generate the topological sequence of directed acyclic graph. The parallel implementation of the proposed GA in CUDA environment consisted of two kernels. The former one was used for making the topological sequences by the help of topological sorting, while the later one was hired to calculate the makespan from a longest path algorithm. The crossover and mutation were performed between two randomly selected schedules on CPU. Experiments was setup with Intel(R) Xeon(R) E5-2650 @ 2.00 GHz and NVIDA Tesla C2075 (448 cores) and results have shown the proposed GA performed around 9 times faster for large-scale problems than the sequential GA.

Another job shop scheduling problem was studied by Mui et al. [17] where a prior-rule was used to create active schedules. The selection combined the idea of an elitist strategy and a roulette wheel selection, whereas the crossover hired a GT algorithm implemented on three parents and the mutation used neighborhood searching technique. With this design, the main part of the GA could be computed independently. In a parallel environment, the master-slave GA was employed where the slaves performed the GA evolutionary operators concurrently and the master searched the global optimum among optimal results received from slaves. The proposed method was run on the CSS computer server system with 6 computers, in which each computer had a Pentium-4 CPU with 4GB free of ram. Empirical results have shown the master-slave GA with 6 processors could save 3 to 4 times the execution time compared to the sequential version.

### 2) flow shop scheduling problems

A master-slave GA dealing with a single population and a group of local subpopulation was presented in [18] for a flow shop problem. This method involved a master scheduler and a set of slave processors. The master scheduler ran the GA operators (partial replacement selection, cycle crossover and swapping mutation) of all individuals sequentially. When the evolution of one individual was finished, it was placed in the unassigned queue from which the master scheduler partitioned the fitness value calculation to slave processors in batches. The choice of candidate slave processors was made upon the involved communication overhead and their computational potential. The available resources among slave processors in the distributed system could vary over time. Moreover, all individuals were maintained in the master scheduler synchronously. The proposed GA was implemented on a laptop with Prentium IV core 2 Duo 2.53 GHz CPU. The outputs showed the new algorithm could be 9 times faster maximally than the results of serial GA achieved by the Lingo 8 software.

Attentions to use master-slave GA to shop scheduling problems are increased in the last decade and the work is carried with various underlying architectures. Since only independent tasks are executed on slaves without communication cost among them, both the conventional GA and any improved GAs can be implemented with it easily. Although the communication between the master and the slaves is an impediment in speed, it still performs well to solve shop scheduling problems whose fitness value calculation is complex and requires considerable computation.

## C. Fine-grained Genetic Algorithms

Fine-grained GA can also be called as neighborhood GA, diffusion GA or massively parallel GA. The main idea is to map individuals of a single GA population on a spatial structure. An individual is limited to compete and mate with its neighbors, while the neighborhoods overlapping makes good solutions disseminate through the entire population. This model obtains good population diversity when dealing with high-dimensional variable spaces [19]. Meanwhile, it is easy to be placed in any two dimensional grid, as many massively parallel machines are designed with this topology. However, we cannot neglect the great influence from the spatial structure, which generally has little chance to be modified. The implemented process of the fine-grained GA is shown in Table 4.

TABLE IV PSEUDO-CODE OF THE FINE-GRAINED GA

| | |
|---|---|
| 1: | Initialize(); |
| 2: | while (termination criteria are not satisfied) do |
| 3: |    Generation++ |
| 4: |    Parallel_NeighborhoodSelection_Individuals(); |
| 5: |    Parallel_NeighborhoodCrossover_Individuals(); |
| 6: |    Parallel_Mutation_Individuals(); |
| 7: |    Parallel_FitnessValueEvaluation_Individuals(); |
| 8: | end while |

*1）job shop scheduling problems*

A fine-grained GA solving job shop scheduling problems was considered by Tamaki et al. [20]. In this paper, the selection was executed locally in a neighborhood of each population. The objectives of this neighborhood model were to improve search in the GA by suppressing favorably the premature convergence phenomena, and to reduce computational time by implementing it on a parallel computer at the same time. The method was then modified as an absolute neighborhood model and implemented on Transputer. Transputer was a MIMD (Multi-Instruction Multi-Data) machine with microprocessors, featuring integrated memory and serial communication links. Through several computational experiments for job shop scheduling problems, the parallel GA with 16 processors could shorten the calculation time dramatically. However, as Transputer did not equip with shared memory, the information exchange was handled through communication operations. Therefore, the calculation time reduction was not able to reach an ideal level. Lin et al. [21] investigated parallel GAs on job shop scheduling problems with a direct solution representation, which encoded the operation starting times. The GA operators were inspired by the G&T algorithm with the random selection, the THX (time horizon exchange) crossover and the THX mutation. Two hybrid models built up by the fine-grained GA with a two-dimensional torus topology and the island GA connected in a ring were discussed in this paper. The first one was an embedding of the fine-grained GA into the island GA, in which each subpopulation on the ring was a torus. The migration on the ring was much less frequent than within the torus. In the second model, the connection topology used in the island GA was one which is typically found in the fine-grained GA, and a relatively large number of nodes were used. The migration frequency kept the same in the island GA. Those two methods were carried on a Sun Ultra 1 which was a family of Sun Microsystems workstations based on the 64-bit Ultra SPARC microprocessor with a single population GA, two island GAs of different subpopulation sizes and one torus fine-grained GA. The execution time comparison was only made between the single population GA and island GAs with the speedup of 4.7 and 18.5 respectively. Regarding to solutions quality, best results were obtained by the hybrid model consisting of island GAs connected in a fine-grained GA style topology by combing the merits from them.

Compared with other two kinds of parallel GA, it seems the implementation of fine-grained GA for shop scheduling problems is rare and outdated, no matter the amount of related papers or the various types of problems treated. With the development of modern computing accelerators with two-dimensional grid environment, like GPU, this implementation has a lot of potential in the near future. Apart form manufacturing systems, the fine-grained GA is also hired for task scheduling problems [22]. It is another type of scheduling problems that focuses on minimizing makespan as well but for a set of tasks to be executed in multiprocessor systems. In this domain, the fine-grained model is treated sometimes as the parallel cellular GA [23].

*D. Island Genetic Algorithms*

The island model is the most famous for the research on parallel GAs. In some papers, it may be called as coarse-grained models, multi-deme models, multi-population models, migration models or distributed models. Unlike previous parallel GAs, this model divides the population into a few relatively large subpopulations. Each of them works as an island and is free to converge towards its own sub-optima. At some points, a migration operator is used to exchange individuals among islands. These configurations make the average population fitness improve faster and mix good local feature efficiently [5]. The main idea of this parallelization is a simple extension of the serial GA while the island model based underlying architecture is easily available. Therefore, the island GA dominates the work on parallel GA for shop scheduling problems. A brief outline about this algorithm is illustrated in Table 5.

TABLE V PSEUDO-CODE OF THE ISLAND GA

| | |
|---|---|
| 1: | Initialize(); |
| 2: | while (termination criteria are not satisfied) do |
| 3: |    Generation++ |
| 4: |    Parallel_SubSelection_Islands(); |
| 5: |    Parallel_SubCrossover_Islands(); |
| 6: |    Parallel_SubMutation_Individuals (); |
| 7: |    Parallel_FitnessValueEvaluation_Individuals(); |
| 8: |    if (generation % migration interval==0) |
| 9: |      Parallel_Migration_Islands(); |
| 10: |    end if |
| 11: | end while |

*1）job shop scheduling problems*

Huang et al. [24] discussed flow shop scheduling problems with fuzzy processing times and fuzzy due dates, where the possibility and necessity measures with exact formulas were adopted to maximize the earliness and tardiness simultaneously. A modified GA was designed to solve the problems with the random keys, the parameterized uniform crossover and the immigration. If Pt was the family of chromosomes in the t-th generation, then |Pt| denoted the population size of Pt. The next generation was made of a% best chromosomes from Pt, b% chromosomes for taking crossovers, and c% chromosomes generated randomly as immigrations, where a+b+c=100. In order to get more efficient convergence, an idea of the longest common substring and rearranging of the chromosomes chosen in the mating pool were also imposed in the algorithm. The full procedure was coded on CUDA by separating the whole population into blocks using the block size of 256 or 128. Circumventing to load the random keys of all chromosomes to global memory, one chromosome was distributed to a block so that all random keys could fit in the shared memory. Although there was no migration among blocks, the idea was organized based on the island GA. In the case of 200 jobs, the numerical simulations on a 2.33 GHz Intel

Core2 Quad desktop computer with 2 GB of RAM, and an NVIDIA GeForce GTX285 graphics card showed that the proposed GA combining with CUDA parallel computation got 19 times speedup. Similarly, Zajicek et al. [25] proposed a homogeneous parallel GA model on the CUDA architecture, where all computations were carried out on the GPU in order to reduce communication between CPU and GPU. The main idea was based on an island GA with a tournament selection, an arithmetic crossover and a Gaussian mutation. Experiments were carried on a system with AMD Phenom II X4 945 3.0 GHz processor and NVIDIA Tesla C1060 GPU. Some instances of the flow shop scheduling problem were solved with speedup from 60 to 120 comparing to the equivalent sequential CPU version.

Park et al. [26] studied a hybrid GA and its parallel version for job shop scheduling problems with an operation-based representation. Concerning the parallel GA, the population was divided into two or four subpopulations. Each subpopulation acted as a single-population GA, where some individuals could migrate from one subpopulation to another at certain intervals. As four population initialization methods, four crossover operators and two selection operators were proposed in this paper, different subpopulations were equipped with different settings to help them evolve independently. Beside, the migration was implemented synchronously with a static ring type connection scheme. Experiments were carried out on a PC with Pentium II 350 and 64MB main memory with MT, ORB and ABZ benchmark problems. The outputs confirmed the island GA improved not only the best solution but also the average solution from results of single GA. Asadzadeh et al. addressed a parallel agent-based GA for a job shop scheduling problem in [27]. Chromosomes of the population, indicating feasible schedules for problem instances were created by the management agent and the execute agent. Afterwards, the management agent divided it into subpopulations with the same size and sent each of them to processor agents. Each processor agent located on a distinct host and executed GA with a roulette wheel selection, a partially matched crossover and a subsequent gene mutation on its subpopulation independently. Different subpopulations communicated by exchanging migrants through the synchronization agent. The number of processor agents was fixed at eight in the experiments. Furthermore, those agents formed a virtual cube amongst themselves and each of them had three neighbors. JADE middleware was used to implement this method, which was a software development framework aimed at developing multi-agent systems. Compared with the serial agent-based GA, the suggested algorithm not only obtained much short schedule lengths, but also had higher convergence speed with large size problems. In [28]. Gu et al. constructed a stochastic job shop scheduling problem by a stochastic expected value model. It was solved by a parallel quantum GA organized by the island model with a hybrid star-shaped topology. The information communication was performed through a penetration migration at the upper level and through a quantum crossover at the lower lever. Besides, the roulette wheel selection, the cycle crossover and the Not Gate mutation were designed as GA operators. Computational tests were run on a PC with a Pentium Processor with clock speed of 1.66 GHZ. On the average, the advised method had a better performance of generating optimal or near-optimal solutions with fast convergence speed than a GA or a quantum GA for large instance problems. Spanous et al. [29] designed a parallel GA for solving job shop scheduling problems with an elitist strategy based selection, a path relinking crossover and a swap mutation. The parallelization was set following the islands paradigm. However, one subpopulation merged with another one once the individuals inside stagnated, where the Hamming distance of more than half individuals were less than a predefined value. The process continued until there was only one subpopulation. Experiments were performed on a commodity workstation with a Pentium IV CPU running at 2 GHz with 1 GB RAM, The results indicated the addressed algorithm managed to attain a comparable performance with five recent approaches.

Regarding to solve shop scheduling problems by the island GA, various research have been done with different architectures. We can discover that the works with GPU pay heavier attention on speedup gained from the island GA. On the opposite, the others consider more the improvement for solutions quality and convergence speed. Few implementations have discussed them simultaneously with a fair comparison. Besides, the island connection topology is varied from different papers with different migration strategies. Some of the designs are carried with respect to the underlying architectures, whereas the others are proposed from supporting theories. However, a completely understanding for the effects of migration is still missing.

*2）flow shop scheduling problems*

Bożejko et al. proposed a parallel GA for flow shop scheduling problems in [30]. The algorithm was based on an island model. To implementations, a Multi-Step Crossover Fusion was used to construct a new individual using the best individuals of different subpopulations and worked with the migration operator to complete the communication between different islands. Tests were performed on 4-processors Sun Enterprise 4x400 MHz under the Solaris 7 operating system, which is a MIMD machine of processors without shared memory. Four crossover operators and four mutation operators were considered as GA operators. The efficiency of the island GA was activated with the combination of three strategies: with the same or different start subpopulations, as independent or cooperative search islands and with the same or different genetic operators. Results turned out the strategy of starting the computation from different subpopulations on every processor with different crossover operators and cooperation was significantly better than others. The improvement of the distance to reference solutions and the improvement of the standard deviation were at the level of 7% and 40% respectively, comparing to the sequential GA. A related work by the same team to minimize the total weighted completion time for the flow shop problem with a special case of a single machine was solved by a similar island GA in [31]. The results noted the 8-processor implementation performed the best.

*3）open shop scheduling problems*

Kokosiński et al. [32] studied an open shop scheduling problem and two greedy heuristics, LPT-Task and LPT-Machine, were proposed for decoding chromosomes represented by permutations with repetitions. The GA operators constituted a 2-elements tournament selection, a linear order crossover and a swap mutation or an invert mutation with constant or variable mutation probabilities. An island GA with migration was applied to the parallel version in which every island sent its best emigrants to all other islands and received immigrants from them. Incoming individuals replaced the chromosomes of host subpopulation randomly. The experimental platform was a PC with Pentium 4 processor (3.06 GHz) and 1 GB RAM. Unfortunately, this parallelization did not reveal obvious advantages in the results. A non-preemptive open shop scheduling problem was discussed by Harmanani et al [33]. Except a feasible solution, a chromosome in this paper included a scratch area through which a ReduceGap operation communicated to GA operators: the crossover and the mutation. A hybrid island GA was hired to organize the parallelization where neighboring islands shared their best chromosomes every $G_N$ generation and all islands broadcasted their best chromosome to all other islands every $L_N$ generations, where $G_N \ll L_N$. Islands were connected through an Ethernet network and used the Message Passing Interface (MPI) on a Beowulf cluster. The experiments were executed on a cluster of five machines that were running Linux and MPI. The outputs presented that the proposed method converged to a good solution quickly before it saturated with a speedup between 2.28 and 2.89 for large instances. A similar work was carried by Ghosn et al. in [34] later.

*4) flexible shop scheduling problems*

Defersha et al. [35] considered an island GA for a flexible flow shop scheduling problem with lot streaming. In this case, the batch of each job was split into certain number of unequal consistent sublots. Each sublot of a job underwent a number of operations in a fixed sequence where each operation could be processed by one of several eligible machines. Three commonly used migration topologies: ring, mesh and fully connected were discussed in this paper with a k-way tournament selection, five kinds of crossover and six kinds of mutation applied by different probabilities. A parallel computation environment composed more than 250 interconnected workstations each having an 8-core Intel Xeon 2.8GHz processor was used for experiments. Test problems were run using up to 48 cores and taking MPI for communication. With all problems considered, there were makespan reductions through the island GA. Meanwhile, empirical studies presented the impact from its different parameters. Regarding to topology, the fully connected one outperformed other two. Three migration policies: random-replace-random, best-replace-random and best-replace-worst were tested. Results showed the island GA was not much sensitive to the change of migration policy while the best-replace-random migration policy performed better slightly. The same authors built a mathematical model for a flexible job shop scheduling problem incorporating sequence-dependent setup time, attached or detached setup time, machine release dates, and time lag requirements in [36]. Like the previous work, the GA operators constituted a k-way tournament selection, three assignment operators and two sequencing operators applied by different probabilities. However, islands were connected with a randomly topology which employed randomly generated migration routes for each communication epoch. The method was tested on a similar experimental platform. Results of medium size problems showed the island GA helped improve the solutions quality while it converged to a better solution within the allowable computational time for large size problems where the single GA failed.

An island GA for flexible flow shop scheduling problems was addressed by Belkadi et al. [37] where genome constituted one assignment chromosome and a sequencing chromosome. The GA was implemented on a biprocessor architecture with a roulette wheel selection, a uniform crossover and a mutation similar to the crossover but operated only on the sequencing scheduling chromosome. Four combinations from two island connected typologies (ring and grid with two dimensions) and two replacement strategies (best and random) were tested. The results noted those two parameters did not have significant influence in the variation of makespan. Regarding to the subpopulation size and its related subpopulation amount, the quality of the solution decreased progressively at the same time as the number of subpopulations increased based on the experiments. However, when the complexity of the problem rose up, this influence reduced. Finally, outputs stated the migration interval was the parameter that had the decision influence to the island GA where the quality of the solution improved gradually with increasing migration frequency. Meanwhile, it did not weaken when enlarging the solutions' searching space. A comparison between the island GA and the sequential GA was also carried in this paper. According to empirical results, the island GA always obtained a smaller makespan while its performance of execution time was only discussed with theoretical values based on two processors. Rashidi et al. [38] studied flexible flow shop scheduling problems with unrelated parallel machines, sequence-dependent setup times and processor blocking to minimize the makespan and the maximum tardiness. Different weights were assigned to two criteria to transform the problem into a single-objective function. The individuals inside one island sought for their own single-objective function, and all islands worked in parallel for Pareto optimal solutions. The paired weights in different islands are different with a small deviation between each successive pairs. After executing the conventional GA operators, a local search step or a Redirect procedure were implemented to further cover the Pareto solutions. The comparison was carried between the island GA without or with a local search step and a Redirect procedure where the later one indicated better performance.

As a combination of a shop scheduling problem and a parallel machine scheduling problem, the complexity of flexible shop scheduling problems is increased. According to previous work, the implementation of parallel GAs to this kind of specific problems is only referred by the island GA. In addition to design the algorithm, some of the papers have considered the influence from migration by the connection topology, the migration rate, the migration interval and the migration strategy. A good cooperation of these parameters could decentralize the searching space and enlarge the diversity level to make a GA have better performance while enjoying accelerations. However, current implementations are still limited. Moreover, most of the works address only the improvement to solutions quality. Experimental results to analyze the speedup gained from the island GA are not sufficient. As the increased complexity will lead to longer execution time, it is interesting to consider GPU to solve related problems whose native topology is suitable for the island GA with thousands parallel computing threads.

IV. DESIGN OF PARALLEL GA BASED ON HIGH PERFORMANCE COMPUTING FRAMEWORKS TO SHOP SCHEDULING PROBLEMS

The preliminary work of parallel GAs for shop scheduling problems is implemented by fine-grained models on distributed memory machines. Although the results are outdated, impressive reduction for execution time has been achieved. As the fine-grained GA is easy to be placed on a spatial structure, to coordinate this design with some modern HPC accelerators with two dimensional grid architecture, such as CUDA, should optimize its performance. Moreover, with new requirements from manufacturing systems in the real life, the complexity of shop scheduling problems is increasing. The two dimensional grid topology could organize a greater amount of threads to work in parallel which is more efficient to help find optimal results of strong NP hard problems with large instances. The other problems from the operation search family solved in this way [39] could be persuasive evidences.

The MIMD machine also works with the island GA at the earlier stage. Soon, it is improved to a parallel computation environment or a computer cluster equipped with multiple processors or multi-core processors. The commonly used parallel processing library MPI is generally hired for information sharing through migration. Meanwhile, GPU is involved with its special memory management to work with this design. As there is no strict underlying architecture limitation to implement the island GA when dealing with shop scheduling problems, the islands connected topology is varied. According to the collected papers, the ring topology is used most frequently. But it is hard to judge which topology performs the best. Besides, the cooperated influence between islands connected topology and other migration parameters cannot be neglected. Fortunately, the average results confirm the implementations of island GAs for shop scheduling problems are able to improve solutions quality and gain speedup with reasonable migration design. As this model dominates not only the work on parallel GAs for shop scheduling problems but parallel GAs for other applications, it still has a lot of potential in the future with the popularity of computing nodes providing multiple processors or multi-core processors.

Since the master slave GA does not assume underlying computer architecture, any parallel computing environment has the chance to use this design without worrying about sharing information. The most time consuming part for GAs to shop scheduling problems is the fitness value calculation that requires even much longer execution time with large problems. Therefore, GPU equipped with more parallel threads is supposed to have better performance among several choices.

V. CONCLUSIONS

As one kind of important problem in combinatorial optimization problems, applying parallel GAs for solving shop scheduling problems have caught heavy attentions since the last few decades. This survey addressed some of the most representative publications in this domain and the reviews were classified by the most common parallel GA categories: master-slave models, fine-grained models and island models. An independent section for hybrid models combining two of the above methods was not set, as the related work was few. Those we have considered in this survey were assigned to one of the three basic models according to their main designs. Most works of parallel GAs to search optimal results for scheduling problems in manufacturing systems are currently managed by the island GA. However, the future of implementing the other two parallel models to this field is promising as well by the development of modern computing accelerators with more parallel threads.

ACKNOWLEDGMENT

This work was supported by a scholarship from the China Scholarship Council (CSC).